# A General Classification of Electrolyte Solutions Based on Their Charge and Mass Transport Properties

B. Roling, J. Kettner, V. Miß

*Department of Chemistry and Center for Materials Science (WZMW), University of Marburg, Hans-Meerwein-Strasse 4, 35032 Marburg, Germany*

(Dated: March 12, 2021)

**Abstract**

The conventional classification of electrolyte solutions as "strong" or "weak" accounts for their charge transport properties, but neglects their mass transport properties, and is not readily applicable to highly concentrated solutions. Here, we use the Onsager transport formalism in combination with linear response theory to attain a more general classification taking into account both charge and mass transport properties. To this end, we define a molar mass transport coefficient $\Lambda_{mass}$, which is related to equilibrium center-of-mass fluctuations of the mobile ions and which is the mass-transport analogue of the molar ionic conductivity $\Lambda_{charge}$. Three classes of electrolyte solution are then distinguished: (i) "Strong electrolytes" with $4 \cdot \Lambda_{mass} \approx \Lambda_{charge}$; (ii) "weak charge transport electrolytes" with $\Lambda_{charge} \ll 4 \cdot \Lambda_{mass}$; and (iii) "weak mass transport electrolytes" with $4 \cdot \Lambda_{mass} \ll \Lambda_{charge}$. While classes (i) and (ii) encompass the classical "strong" and "weak" electrolytes, respectively, many highly concentrated electrolytes fall into class (iii) and thus exhibit transport properties clearly distinct from classical strong and weak electrolytes.

Electrolyte solutions play a very important role in many scientific fields, such as electrochemistry, energy research, chemical synthesis, biochemistry and pharmaceutical research.[1–10] The conventional classification of electrolyte solutions as "strong" or "weak" is based on the dissociation equilibrium $AX_{(solv)} \rightleftharpoons A^+_{(solv)} + X^-_{(solv)}$. Here, $AX_{(solv)}$ denotes solvated ion pairs, while $A^+_{(solv)}$ and $X^-_{(solv)}$ denote solvated free ions contributing to the ionic conductivity of the solution. The degree of dissociation $\alpha = c_{A^+}/(c_{A^+} + c_{AX}) = c_{X^-}/(c_{X^-} + c_{AX})$ is large ($\alpha \approx 1$) in the case of strong electrolytes and small ($\alpha \ll 1$) in the case of weak electrolytes. The molar ionic conductivity of an electrolyte solution, $\Lambda_{charge} = \sigma_{ion}/c_{salt}$, with $\sigma_{ion}$ and $c_{salt} = c_{A^+} + c_{AX}$ denoting the ionic conductivity and the overall salt concentration, respectively, increases with increasing degree of dissociation $\alpha$. Thus, weak electrolytes, such as aqueous solutions of acetic acid, exhibit much lower molar ionic conductivities than strong electrolytes, such as aqueous KCl solutions. On the other hand, the mass transport properties of strong and weak electrolytes are similar, since free ions and solvated ion pairs exhibit typically similar self diffusion coefficients and thus contribute both to the mass transport. For instance, the chemical diffusion coefficient of an acetic acid solution depends very weakly on the acetic acid concentration and thus on the degree of dissociation.[11] It is important to emphasize that for electrochemical applications of electrolyte solutions, both charge and mass transport properties are highly relevant. As an example, in alkali-ion batteries, the transport of alkali ions between the electrodes takes place via a combination of electric field-induced migration and salt concentration gradient-induced diffusion.[12,13] Consequently, an electrolyte classification taking into account both charge and mass transport properties is highly desirable.

Currently, there is huge interest in using highly concentrated electrolytes for electrochemical energy storage in batteries and supercapacitors.[14–17] In highly concentrated electrolytes, a large part or even a major part of the solvent molecules is involved in the solvation of the ions. This results in a low chemical potential of the solvent molecules, leading to a low vapor pressure and to a broad electrochemical stability window of the electrolyte solution. An interesting example are water-in-salt electrolytes[18], which exhibit a much broader electrochemical stability window than diluted aqueous solutions and thus pave the way for the development of aqueous Li-ion batteries.[19] From a basic science point of view, there are controversial viewpoints as to whether highly concentrated electrolytes exhibit a low degree of dissociation ("low ionicity") and can thus be classified as "weak" or "diluted" electrolytes.[20–25]

In this Letter, we show that a more general classification of electrolyte solutions, encompassing also highly concentrated solutions, can be achieved by taking into account both charge and mass transport properties. We use the Onsager transport formalism in combination with linear response theory to define a molar mass transport coefficient, $\Lambda_{mass}$, which is the mass-transport analogue of the molar ionic conductivity, $\Lambda_{charge}$. We show that $\Lambda_{mass}$ is related to equilibrium center-of-mass fluctuations caused by the ion movements, while it is well known that $\Lambda_{ion}$ is related to equilibrium dipole fluctuations caused by the ion movements. By analyzing the relations between $\Lambda_{mass}$ and $\Lambda_{charge}$, we show that three classes of electrolyte solutions can be distinguished. In the following, we start with the theoretical background and then analyze transport data for a number of 1-1 electrolyte solutions (univalent cations and univalent anions) taken from experiment and molecular dynamics simulations.

In Pfeifer et al. [26], it was shown that the Onsager transport coefficients $\sigma_{++}$, $\sigma_{--}$, and $\sigma_{+-}$ of a 1-1 electrolyte can be expressed as:

$$\sigma_{++} = 2 \cdot \sigma_0 \cdot \left( \langle A^2 \rangle + \frac{k^2}{(1+k)^2} \langle B^2 \rangle \right) \tag{1a}$$

$$\sigma_{--} = 2 \cdot \sigma_0 \cdot \left( \langle A^2 \rangle + \frac{1}{(1+k)^2} \langle B^2 \rangle \right) \tag{1b}$$

$$\sigma_{+-} = 2 \cdot \sigma_0 \cdot \left( \langle A^2 \rangle - \frac{k}{(1+k)^2} \langle B^2 \rangle \right) \tag{1c}$$

The prefactor $\sigma_0$ is proportional to the overall number density of cations and anions [26] and thus to the overall salt concentration $c_{salt}$. $\langle A^2 \rangle$ and $\langle B^2 \rangle$ characterize equilibrium center-of-mass fluctuations and equilibrium dipole fluctuations, respectively, caused by the ion movements, while $k = \frac{m_-}{m_+}$ denotes the mass ratio of the ions.[26] The overall ionic conductivity is then given by $\sigma_{ion} = \sigma_{++} + \sigma_{--} - 2\sigma_{+-} = 2 \cdot \sigma_0 \langle B^2 \rangle$, yielding the following expression for the molar ionic conductivity:

$$\Lambda_{charge} = \frac{\sigma_{ion}}{c_{salt}} = \frac{2 \cdot \sigma_0 \langle B^2 \rangle}{c_{salt}} \tag{2}$$

Since $\sigma_0$ is proportional to $c_{salt}$, the ratio $\frac{\sigma_0}{c_{salt}}$ is independent of $c_{salt}$, and thus $\Lambda_{charge}$ is determined by the mean squared dipole fluctuations caused by the ion movements in equilibrium.

Now we define a mass transport coefficient as follows:

$$\sigma_{mass} = \frac{\sigma_{++} \cdot \sigma_{--} - (\sigma_{+-})^2}{\sigma_{ion}} \tag{3}$$

As shown in the Appendix, $\sigma_{mass}$ relates the salt flux to the chemical potential gradient of the salt. Inserting Eqs. (1a)-(1c) into Eq. (3) yields:

$$\sigma_{mass} = 2 \cdot \sigma_0 \cdot \langle A^2 \rangle \tag{4}$$

Dividing by the salt concentrations $c_{salt}$ results in the following expression for a molar mass transport coefficient $\Lambda_{mass}$:

$$\Lambda_{mass} = \frac{\sigma_{mass}}{c_{salt}} = \frac{2 \cdot \sigma_0 \langle A^2 \rangle}{c_{salt}} \tag{5}$$

Since $\frac{\sigma_0}{c_{salt}}$ is independent of $c_{salt}$, $\Lambda_{mass}$ is proportional to the mean squared center-of-mass fluctuations caused by the ion movements in equilibrium, and is the mass transport analogue of $\Lambda_{charge}$.

Now we calculate $\Lambda_{charge}$ and $\Lambda_{mass}$ for a classical "weak" electrolyte, namely acetic acid (HOAc) solutions in water. For solutions with concentrations up to 0.1 mol/dm³, thermodynamic activity coefficients and thermodynamic factors are very close to unity.[11] In this case, the degree of dissociation $\alpha$ can be directly calculated from the concentration-based equilibrium constant $K_c = \frac{c_{H^+} \cdot c_{OAc^-}}{c_{HOAc} \cdot c^\ominus}$ and the overall salt concentration $c_{salt}$ via $\alpha = \frac{K_c \cdot c^\ominus}{2 \cdot c_{salt}} \cdot \left[ \sqrt{1 + \frac{4 \cdot c_{salt}}{K_c \cdot c^\ominus}} - 1 \right]$. Here, $c^\ominus$ denotes the standard concentration 1 mol/dm³. The Onsager coefficients are then given by [13]:

$$\sigma_{++} = \frac{c_{salt} F^2}{RT} D^*_{H^+} \tag{6a}$$

$$\sigma_{--} = \frac{c_{salt} F^2}{RT} D^*_{OAc^-} \tag{6b}$$

$$\sigma_{+-} = \frac{(1-\alpha) c_{salt} F^2}{RT} D^*_{HOAc} \tag{6c}$$

with $D^*_{H^+}$, $D^*_{OAc^-}$, and $D^*_{HOAc}$ denoting the self-diffusion coefficients of free H⁺ ions, of free OAc⁻ ions and of neutral HOAc molecules, respectively. For acetic acid, we have $K_c = 1.7 \cdot 10^{-5}$, $D^*_{H^+} = 9.3 \cdot 10^{-5}$ cm²/s, $D^*_{OAc^-} = 1.1 \cdot 10^{-5}$ cm²/s, and $D^*_{HOAc} = 1.2 \cdot 10^{-5}$ cm²/s.[11] Based on these values, we calculate $\Lambda_{charge}$ and $\Lambda_{mass}$ for HOAc concentrations in the range from $10^{-3}$ mol/dm³ to $10^{-1}$ mol/dm³, in which the degree of dissociation is $\alpha \ll 1$, and we plot $\Lambda_{mass}$ vs. $\Lambda_{charge}$, see Fig. 1. For comparison, we have added a line for an ideal strong electrolyte with complete ion dissociation ($\alpha = 1$), with negligible ion-ion interactions and with identical masses of cations and anions ($k = 1$). Negligible ion-ion interactions implies that $\sigma_{+-} = 0$.[13] From Eq. (1c), (2), and (5) it follows that in the ideal strong electrolyte case:

$$\frac{\langle A^2 \rangle}{\langle B^2 \rangle} = \frac{k}{(1+k)^2} = \frac{\Lambda_{mass}}{\Lambda_{charge}} \tag{7}$$

Thus, in the case of $k = 1$, we have $4 \cdot \Lambda_{mass} = \Lambda_{charge}$. Note that $\frac{k}{(1+k)^2}$ does not deviate strongly from ¼, when the ion mass ratio varies between $\frac{1}{2}$ and 2.

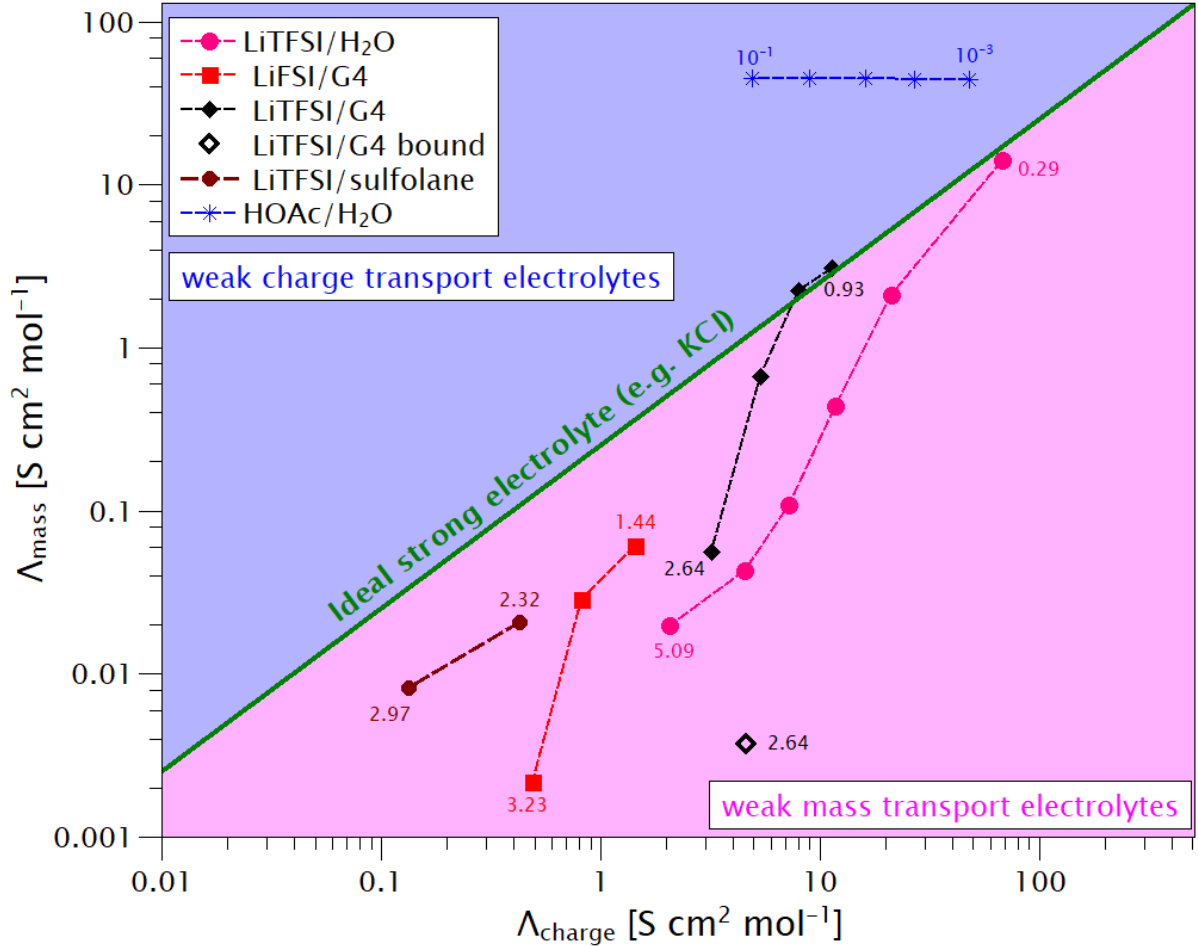

*Fig. 1: Plot of the molar mass transport coefficient $\Lambda_{mass}$ vs. the molar ionic conductivity $\Lambda_{charge}$ for a number of 1-1 electrolyte solutions. The green line refers to the transport properties of an ideal strong electrolyte (no ion-ion interactions) with identical masses of cations and anions. The numbers added to some data points refer to the salt concentration in the respective solutions.*

The data points for the HOAc solutions are characterized by $4 \cdot \Lambda_{mass} > \Lambda_{charge}$ and are thus located above the ideal strong-electrolyte line, see Fig. 1.

Next, we use Onsager coefficients for semiconcentrated and highly concentrated battery electrolyte solutions from the literature for calculating $\Lambda_{charge}$ and $\Lambda_{mass}$, see Fig. 1. The data for LiFSI / tetraglyme (G4) solutions[26] and for LiTFSI / sulfolane solutions[27] were obtained in experimental studies, while the data for LiTFSI / G4 solutions[28] and LiTFSI / H$_2$O solutions[29] were derived from molecular dynamics simulations. For salt concentrations up to about 1 mol/dm$^3$, the $\Lambda_{mass}$ vs. $\Lambda_{charge}$ data are located close to the ideal strong electrolyte line, but for high salt concentrations > 2 mol/dm$^3$, the data are located well below the ideal strong electrolyte line. This implies that mass transport in these highly concentrated solutions is much slower than charge transport. The reason for this is the lack of free solvent molecules in these solutions. In equilibrium, the center of mass of the entire electrolyte solution cannot fluctuate. Therefore, center-of-mass fluctuations of the ions are only possible via momentum exchange with free solvent molecules. If the number of free solvent molecules is small, center-of-mass fluctuations of the ions are weaker than dipole fluctuations, and according to Eqs. (2) and (5), mass transport is slower than charge transport.

The lack of free solvent molecules is particular pronounced in equimolar mixtures of Li salts and tetraglyme (G4) due to the strong binding between Li$^+$ ions and G4 molecules. [26,28] The strongest effect regarding $4 \cdot \Lambda_{mass} \ll \Lambda_{charge}$ is found in an MD simulation of an equimolar mixture of LiTFSI and G4 with harmonic (i.e. unbreakable) bonds between Li$^+$ ions and G4 molecules.[28] The effect is weaker for water and sulfolan as solvents, since the bonds between Li$^+$ ions and these solvent molecules are weaker, so that there is a faster exchange of solvent molecules between the Li$^+$ ions.[29] This facilitates center-of-mass fluctuations of the ions.

From these results, it follows that three classes of electrolyte solutions can be distinguished: (i) Strong electrolytes are characterized by $\frac{\Lambda_{mass}}{\Lambda_{charge}} = \frac{k}{(1+k)^2} \approx \frac{1}{4}$. (ii) Classical weak electrolytes, like aqueous HOAc solutions, are characterized by $\Lambda_{charge} \ll 4 \cdot \Lambda_{mass}$. Consequently, we suggest to use the term "weak charge transport electrolytes" for this class. (iii) Many concentrated electrolyte solutions are characterized by $4 \cdot \Lambda_{mass} \ll \Lambda_{charge}$, and we suggest to use the term "weak mass transport electrolytes" for this class.

In order to support this classification of electrolyte solutions, the typical Walden plot of $\Lambda_{charge}$ vs. reciprocal viscosity $\eta^{-1}$ can be complemented by a plot of $4 \cdot \Lambda_{mass}$ vs. $\eta^{-1}$. In Fig. 2, such Walden plots are shown exemplarily for two solutions, for which viscosity data are available, namely for HOAc/H$_2$O solutions [30] and for LiTFSI/H$_2$O solutions.[29] The data show clearly that these solutions belong to different classes. The $\Lambda_{charge}$ vs. $\eta^{-1}$ data of the HOAc/H$_2$O solutions are located well below the ideal KCl line, while the $4 \cdot \Lambda_{mass}$ vs. $\eta^{-1}$ data are located close to the ideal line (even slightly above the ideal line). The data of the LiTFSI/H$_2$O solutions show the opposite trend, i.e. the $\Lambda_{charge}$ vs. $\eta^{-1}$ data are located close to the ideal KCl line, while the $4 \cdot \Lambda_{mass}$ vs. $\eta^{-1}$ data at high concentrations are located well below the ideal line.

In conclusion, we have shown that the conventional classification of electrolyte solutions in terms of "strong" and "weak" does not account for their mass transport properties and is therefore not readily applicable to highly concentrated electrolyte solutions. The theoretical background and the data analysis presented in this paper demonstrates that three classes of electrolyte solutions should be distinguished, namely (i) "strong electrolytes" with similar values of $\Lambda_{charge}$ and $4 \cdot \Lambda_{mass}$, (ii) "weak charge transport electrolytes" with $\Lambda_{charge} \ll 4 \cdot \Lambda_{mass}$, and (iii) "weak mass transport electrolytes" with $4 \cdot \Lambda_{mass} \ll \Lambda_{charge}$. The distinction between these classes is essential for a comprehensive understanding of the transport properties of electrolyte solutions in electrochemical devices.

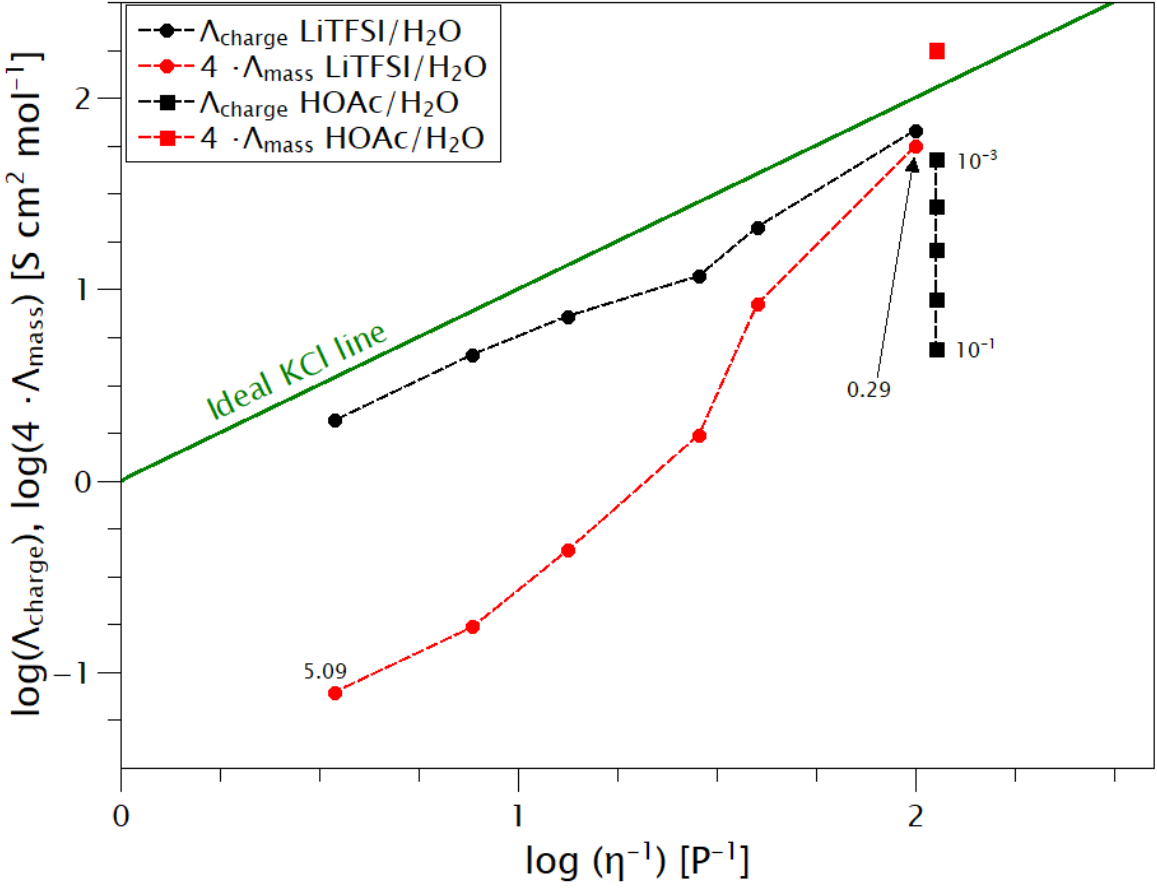

*Fig. 2: Walden plot of the molar mass transport coefficient $\Lambda_{mass}$ and of the molar ionic conductivity $\Lambda_{charge}$, respectively, versus the reciprocal viscosity of the electrolyte solution. The numbers added to some data points refer to the salt concentration in the respective solutions.*

**Acknowledgements**

We would like to thank the Federal State of Hesse (Germany) for financial support of this work.

**Appendix: Derivation of Eq. (3)**

In the Onsager transport formalism, the molar fluxes of cations and anions, $J_+$ and $J_-$, are proportional to their electrochemical potential gradients $\nabla\tilde{\mu}_+$ and $\nabla\tilde{\mu}_-$. For an electrolyte solution with one type of univalent cations and one type of univalent anions, we can write:

$$\boldsymbol{J}_+ = -\left(\frac{\sigma_{++}}{F^2}\right)\nabla\tilde{\boldsymbol{\mu}}_+ - \left(\frac{\sigma_{+-}}{F^2}\right)\nabla\tilde{\boldsymbol{\mu}}_- \tag{A1a}$$

$$\boldsymbol{J}_- = -\left(\frac{\sigma_{--}}{F^2}\right)\nabla\tilde{\boldsymbol{\mu}}_- - \left(\frac{\sigma_{+-}}{F^2}\right)\nabla\tilde{\boldsymbol{\mu}}_+ \tag{A1b}$$

$\sigma_{ij}$ denote Onsager transport coefficients, with the cross coefficient $\sigma_{+-}$ accounting for interactions between cations and anions. $F$ is the Faraday constant.

Now we consider an electrochemical potential gradient acting only in one dimension $x$, and we split the electrochemical potential gradient into an electrical potential term and a chemical potential gradient term:

$$J_+ = -\left[\left(\frac{\sigma_{++}}{F^2}\right)\left(F\frac{d\varphi}{dx}+\frac{d\mu_+}{dx}\right)+\left(\frac{\sigma_{+-}}{F^2}\right)\left(-F\frac{d\varphi}{dx}+\frac{d\mu_-}{dx}\right)\right] \tag{A2a}$$

$$J_- = -\left[\left(\frac{\sigma_{--}}{F^2}\right)\left(-F\frac{d\varphi}{dx}+\frac{d\mu_-}{dx}\right)+\left(\frac{\sigma_{+-}}{F^2}\right)\left(F\frac{d\varphi}{dx}+\frac{d\mu_+}{dx}\right)\right] \tag{A2b}$$

Here, $\mu_+$ and $\mu_-$ are the chemical potentials of cations and anions, respectively, while $\varphi$ denotes the electrical potential. It is usually assumed that the chemical potentials of cations and anion cannot be determined individually. Consequently, a mean activity coefficient $\gamma_{+-}$ is defined such that:

$$d\mu_+ = RT \cdot d\ln a_+ = RT \cdot d\ln(\gamma_{+-} \cdot c_+) \tag{A3a}$$

$$d\mu_- = RT \cdot d\ln a_- = RT \cdot d\ln(\gamma_{+-} \cdot c_-) \tag{A3b}$$

Here, $R$ and $T$ denote the ideal gas constant and temperature, respectively, while $a_+$ and $a_-$ are the activities of cations and anions, respectively. With $d\mu_+ + d\mu_- = d\mu_{salt}$ and $c_+ = c_-$ (electroneutrality) it follows that

$$d\mu_+ = d\mu_- = \frac{1}{2}d\mu_{salt} \tag{A4}$$

Neutral salt diffusion implies that $J_+ = J_- = J_{salt}$. With Eqs. (A2a), (A2b) and (A4) this leads to:

$$F\frac{d\varphi}{dx} = \left(\frac{-\sigma_{++}+\sigma_{--}}{\sigma_{++}+\sigma_{--}-2\sigma_{+-}}\right)\frac{1}{2}\frac{d\mu_{salt}}{dx} \tag{A5}$$

Solving Equation (A5) for $d\varphi/dx$ and inserting the result into Equation (A2a) gives:

$$\begin{aligned} J_{salt} &= -\left[\left(\frac{\sigma_{++}-\sigma_{+-}}{F}\right)\left[\left(\frac{-\sigma_{++}+\sigma_{--}}{\sigma_{++}+\sigma_{--}-2\sigma_{+-}}\right)\right]\frac{1}{2}\frac{d\mu_{salt}}{dx}+\left(\frac{\sigma_{++}+\sigma_{+-}}{F}\right)\frac{1}{2}\frac{d\mu_{salt}}{dx}\right]\\ &= -\frac{1}{F}\left[\frac{\sigma_{++}\sigma_{--}-\sigma_{+-}^2}{\sigma_{++}+\sigma_{--}-2\sigma_{+-}}\right]\frac{d\mu_{salt}}{dx}\\ &= -\frac{1}{F}\cdot\sigma_{mass}\frac{d\mu_{salt}}{dx} \end{aligned} \tag{A6}$$